# MIDI-LAB, a Powerful Visual Basic Program for Creating MIDI Music

Kai Yang [1], Xi Zhou [2]

[1] School of Computing, University of Utah, Utah, USA
kai.yang@utah.edu
[2] Department of Family & Preventive Medicine, University of Utah, Utah, USA
xi.zhou@utah.edu

## *Abstract*

*Creating MIDI music can be a practical challenge. In the past, working with it was difficult and frustrating to all but the most accomplished and determined. Now, however, we are offering a powerful Visual Basic program called MIDI-LAB, that is easy to learn, and instantly rewarding to even the newest users. MIDI-LAB has been developed to give users the ability to quickly create music with a limitless variety of tunes, tempos, speeds, volumes, instruments, rhythms and major scales. This program has a simple, intuitive, and user-friendly interface, which provides a straightforward way to enter musical data with Numbered Musical Notation (NMN) and immediately create MIDI music. The key feature of this program is the digitalization of music input. It vastly simplifies creating, editing, and saving MIDI music. MIDI-LAB can be used virtually anywhere to write music for entertainment, teaching, computer games, and mobile phone ringtones.*

## 1. Instruction

MIDI (Musical Instrument Digital Interface) is a digital communications language and compatible hardware specification that allows multiple electronic instruments, performance controllers, computers, and other related devices to communicate with each other within a connected network [1]. Standard MIDI files are a popular source of music, and MIDI music is widely accepted by musicians, composers, music lovers, teachers, mobile phone users, and game makers. Especially, MIDI has demonstrated a very broad application prospects in music education [2]. Some scholars in Finland have developed a MIDI Toolbox for analyzing and visualizing MIDI files [3], however, it can only work in Matlab computing environment. Of course, there are several notable software MIDI editors and sequencers on the market, such as Adobe Audition, Anvil Studio, Cakewalk, Cubase, and Sekaiju [4]. While these software programs are useful, their complex interfaces are tedious, and take a long time to learn. What's more, most of them are designed like a Piano Roll, i.e. the time axis is located horizontally, and different positions on the vertical line correspond to different pitches. This interface makes it time consuming to read, edit, copy, cut, and paste the tunes and tempos, especially for new users, novices, children, and individuals with dexterity challenges.

Developing a more serviceable, visual, and intuitive MIDI program was imperative. In this paper, we propose a unique Visual Basic program as illustrated in **Figure 1**. This concise interface can be divided into 3 sections, which are the editing, adjusting, and operating sections. The Editing Section contains a Tune Box and a Tempo Box. Apparently, every melody is composed of a series of notes, and the two most important attributes of each note are frequency and duration. Therefore, we can input a sequence of tune numbers in the Tune Box and a sequence of tempo numbers in the Tempo Box to identify a melody. The parameters in the Adjusting Section, such as speed, volume, instrument, rhythm, and major scale are used to change the character of the music. The four buttons in the Operation Section are used to accomplish four basic operations that are play, stop, clear the input boxes, and close the program. The key feature of this program is that most of the input parameters, such as tune, tempo, speed and volume, are digitalized. This key feature makes the music input much simpler to learn and use.

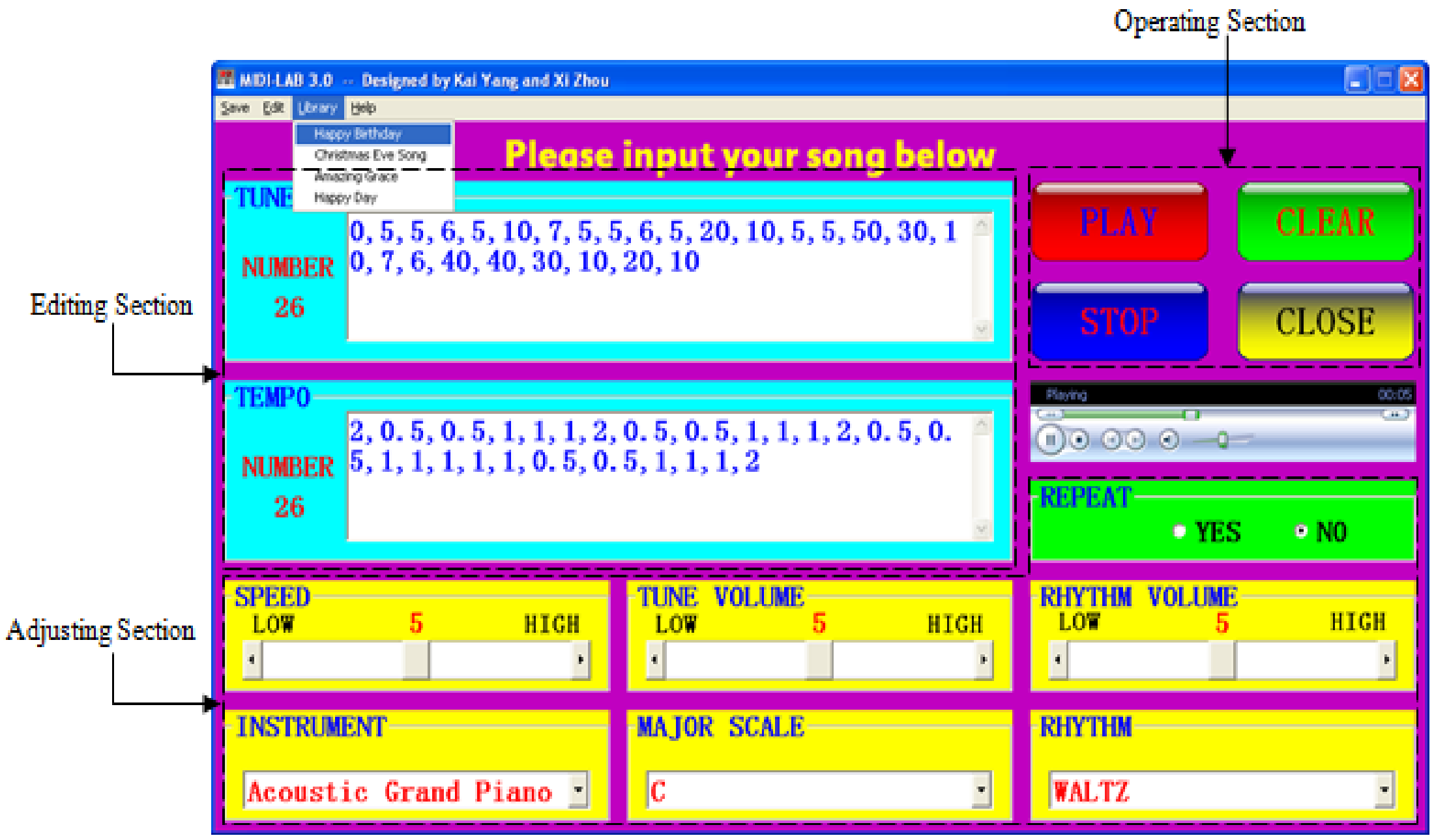

Figure 1. The interface of MIDI-LAB.

## 2. Numbered Musical Notation

Numbered Musical Notation (NMN) is a musical notation system invented by Jean-Jacques Rousseau (1712-1778), a French philosopher of the Enlightenment [5]. NMN takes full advantage of the succinctness of numbers, thereby simplifying traditional Staff Musical Notation (SMN). The integers 1 to 7 represent the seven notes of the diatonic major scale, and number 0 represents the musical rest, which is an interval of silence in a piece of music. Dots above a note indicate octaves higher, and dots below indicate octaves lower. For instance, “5” means Note G, “$\dot{5}$” means an octave higher than Note G, “$\ddot{5}$” means two octaves higher than Note G, “$\underset{\cdot}{5}$” means an octave lower than Note G, and “$\underset{:}{5}$” means two octaves lower than Note G. The underlining of a note or a musical rest shortens it, while dots and dashes after lengthen it. The SMN and NMN of “Happy Birthday to You” melody are shown in **Figure 2**. Compared with SMN, NMN is very compact for just the melody line or monophonic parts. This feature makes the tunes and tempos easy to read, edit, copy, cut, and paste. Hence, MIDI-LAB chooses NMN over the time-consuming SMN.

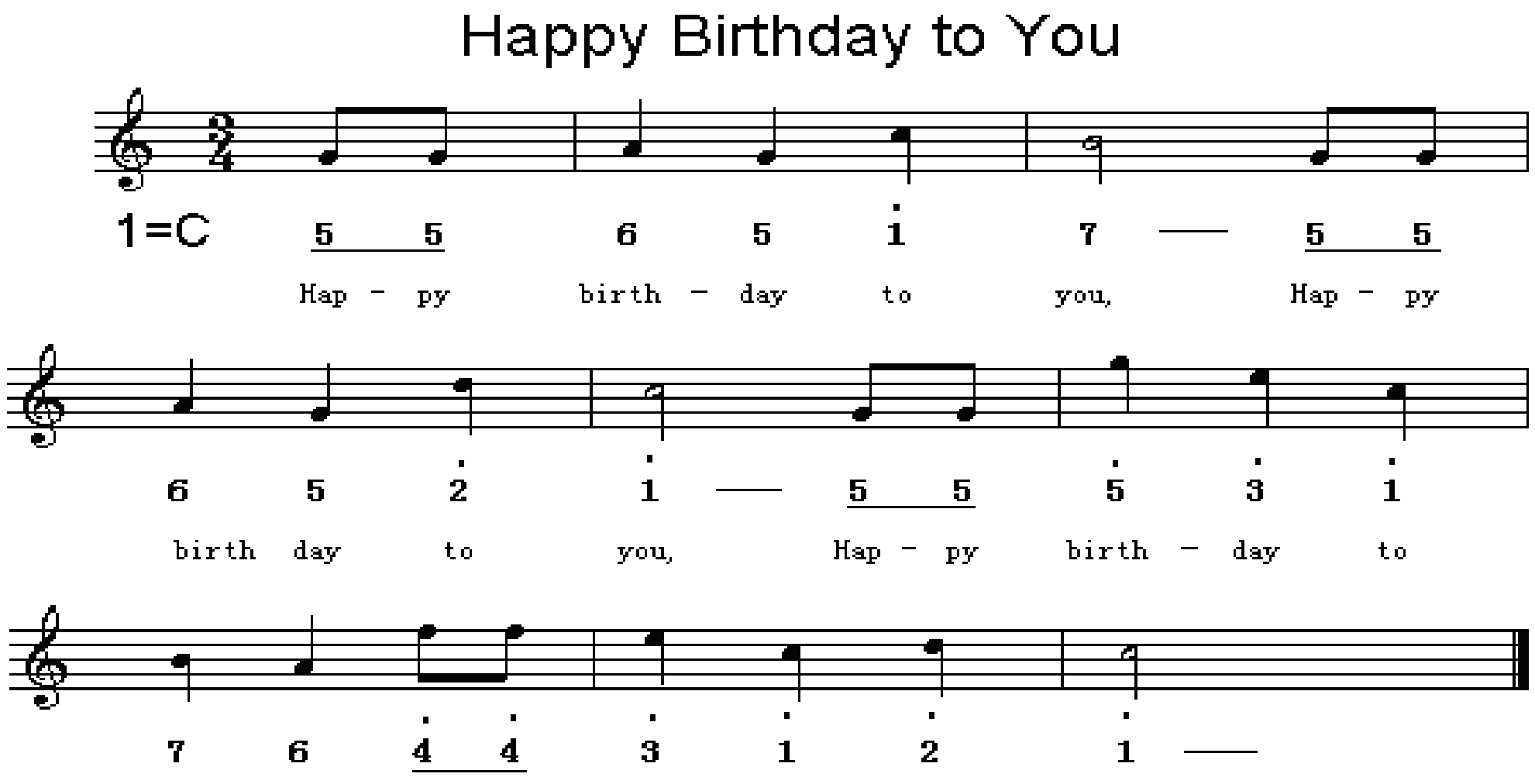

Figure 2**.** SMN and NMN of “Happy Birthday to You” melody.

## 3. Standard MIDI file

One reason for the popularity of MIDI files (.MID) is that they do not actually record the sound of the keyboard instrument [6]. Instead, MIDI files store music data as event messages, This way, MIDI files are much smaller than recorded audio waveforms, and hence require much less storage space. A MIDI file always starts with a header chunk, and is followed by one or more track chunks [7]. The header chunk consists of a literal string denoting the header, a length indicator, the format of the MIDI file, the number of tracks in the file, and a timing value specifying delta time units. A track chunk consists of a literal identifier string, a length indicator specifying the size of the track, and actual event data making up the track. **Table 1** is an example of the hex code of “Happy Birthday to You” melody in MIDI format.

Table 1 "Happy Birthday to You" melody in MIDI format

| | | | | | | | | | | | | | | | |
|---|---|---|---|---|---|---|---|---|---|---|---|---|---|---|---|
| 4D | 54 | 68 | 64 | 00 | 00 | 00 | 06 | 00 | 00 | 00 | 01 | 00 | 03 | 4D | 54 |
| 72 | 6B | 00 | 00 | 00 | 6F | 00 | C0 | 00 | 00 | 90 | 43 | 64 | 02 | 90 | 43 |
| 64 | 02 | 90 | 45 | 64 | 04 | 90 | 43 | 64 | 04 | 90 | 48 | 64 | 04 | 90 | 47 |
| 64 | 08 | 90 | 43 | 64 | 02 | 90 | 43 | 64 | 02 | 90 | 45 | 64 | 04 | 90 | 43 |
| 64 | 04 | 90 | 4A | 64 | 04 | 90 | 48 | 64 | 08 | 90 | 43 | 64 | 02 | 90 | 43 |
| 64 | 02 | 90 | 4F | 64 | 04 | 90 | 4C | 64 | 04 | 90 | 48 | 64 | 04 | 90 | 47 |
| 64 | 04 | 90 | 45 | 64 | 04 | 90 | 4D | 64 | 02 | 90 | 4D | 64 | 02 | 90 | 4C |
| 64 | 04 | 90 | 48 | 64 | 04 | 90 | 4A | 64 | 04 | 90 | 48 | 64 | 08 | B0 | 7B |
| 00 | 00 | FF | 2F | 00 | | | | | | | | | | | |

Table 2 The Note Numbers corresponding to the NMN

| Octave # | Note Numbers | | | | | | | | | | |
|---|---|---|---|---|---|---|---|---|---|---|---|
| | 1 | 1# | 2 | 2# | 3 | 4 | 4# | 5 | 5# | 6 | 6# | 7 |
| 0 | 0 | 1 | 2 | 3 | 4 | 5 | 6 | 7 | 8 | 9 | 10 | 11 |
| 1 | 12 | 13 | 14 | 15 | 16 | 17 | 18 | 19 | 20 | 21 | 22 | 23 |
| 2 | 24 | 25 | 26 | 27 | 28 | 29 | 30 | 31 | 32 | 33 | 34 | 35 |
| 3 | 36 | 37 | 38 | 39 | 40 | 41 | 42 | 43 | 44 | 45 | 46 | 47 |
| 4 | 48 | 49 | 50 | 51 | 52 | 53 | 54 | 55 | 56 | 57 | 58 | 59 |
| 5 | 60 | 61 | 62 | 63 | 64 | 65 | 66 | 67 | 68 | 69 | 70 | 71 |
| 6 | 72 | 73 | 74 | 75 | 76 | 77 | 78 | 79 | 80 | 81 | 82 | 83 |
| 7 | 84 | 85 | 86 | 87 | 88 | 89 | 90 | 91 | 92 | 93 | 94 | 95 |
| 8 | 96 | 97 | 98 | 99 | 100 | 101 | 102 | 103 | 104 | 105 | 106 | 107 |
| 9 | 108 | 109 | 110 | 111 | 112 | 113 | 114 | 115 | 116 | 117 | 118 | 119 |
| 10 | 120 | 121 | 122 | 123 | 124 | 125 | 126 | 127 | | | | |

Beginning in the upper left corner of **Table 1** and reading from left to right, we find that the hex code of a MIDI file can be divided into several sections. Each section will be explained below.

(A) "4D 54 68 64" designates that this is a MIDI file.
(B) "00 00 00 06" is the four bytes indicating how big the rest of the MIDI Header (C, D, E) is. It's always "00 00 00 06" for Standard MIDI Files (SMF).
(C) "00 00" means that this a Type-0 MIDI file. In other words, it contains a single multi-channel track.
(D) "00 01" means that this MIDI file has only one track.
(E) "00 03" controls the speed of the music. The hexadecimal value "03" means 3 ticks per quarter note.
(F) "4D 54 72 6B" marks the start of the track trunk data. This is where the actual song is stored.
(G) "00 00 00 6F" means the remaining bytes in the track trunk, which in this case is 111 (decimal) bytes.
(H) "00 C0 00" defines the musical instrument number. There are 128 choices of instrument, e.g.

piano, organ, guitar, bass, and brass.

(I) “00 90 43 64” defines the first note. “00” means waiting for 0 time units. “90” is the “Note On” message. “43” is Note Number of 5 (NMN). “64” defines the volume. The Note Numbers corresponding to the NMN are listed in **Table 2**. From **Table 2**, we find that the Note Number ranges from 0 to 127, and can cover the notes in 11 octaves (from $\underset{\vdots}{1}$ to $\overset{\vdots}{5}$).

(J) “02 90 43 64” defines the second note. “02” means waiting for 2 time units, so that the previous note lasts for 2 time units. “90” is the “Note On” message. “43” is Note Number of 5 (NMN). “64” defines the volume.

(K) “02 90 45 64” defines the third note. “02” means waiting for 2 time units, so that the last note lasts for 2 time units. “90” is the “Note On” message. “45” is Note Number of 6 (NMN). “64” defines the volume. The next 22 notes are defined in the same way.

(L) “08 B0 7B 00” means all notes off.

(M) “00 FF 2F 00” shows that the end of the track has been reached.

## 4. Visual Basic program design

This program provides a scientific, visual, and easy-to-use MIDI creator that can be used by musicians, composers, music lovers, teachers, mobile phone users, and game makers in many user fields and will provide an attractive product in many markets. As is well known, Visual Basic (VB) is an ideal programming language for developing sophisticated professional applications for Microsoft Windows. It makes use of the Graphical User Interface for creating robust and powerful applications. In this work, we chose Visual Basic 6.0 to build, test, and compile a new powerful and unparalleled music-enabling program, MIDI-LAB. This program shows strong object-oriented technology, and may run as an executable file in any WINDOWS environment even without the presence of Visual Basic.

A main requirement in developing the Visual Basic program is to make a flowchart of the entire process as shown in **Figure 3**. The flowchart can be mainly divided to three parts, namely data input, data check and data output. This flowchart is followed by a clear explanation of each part.

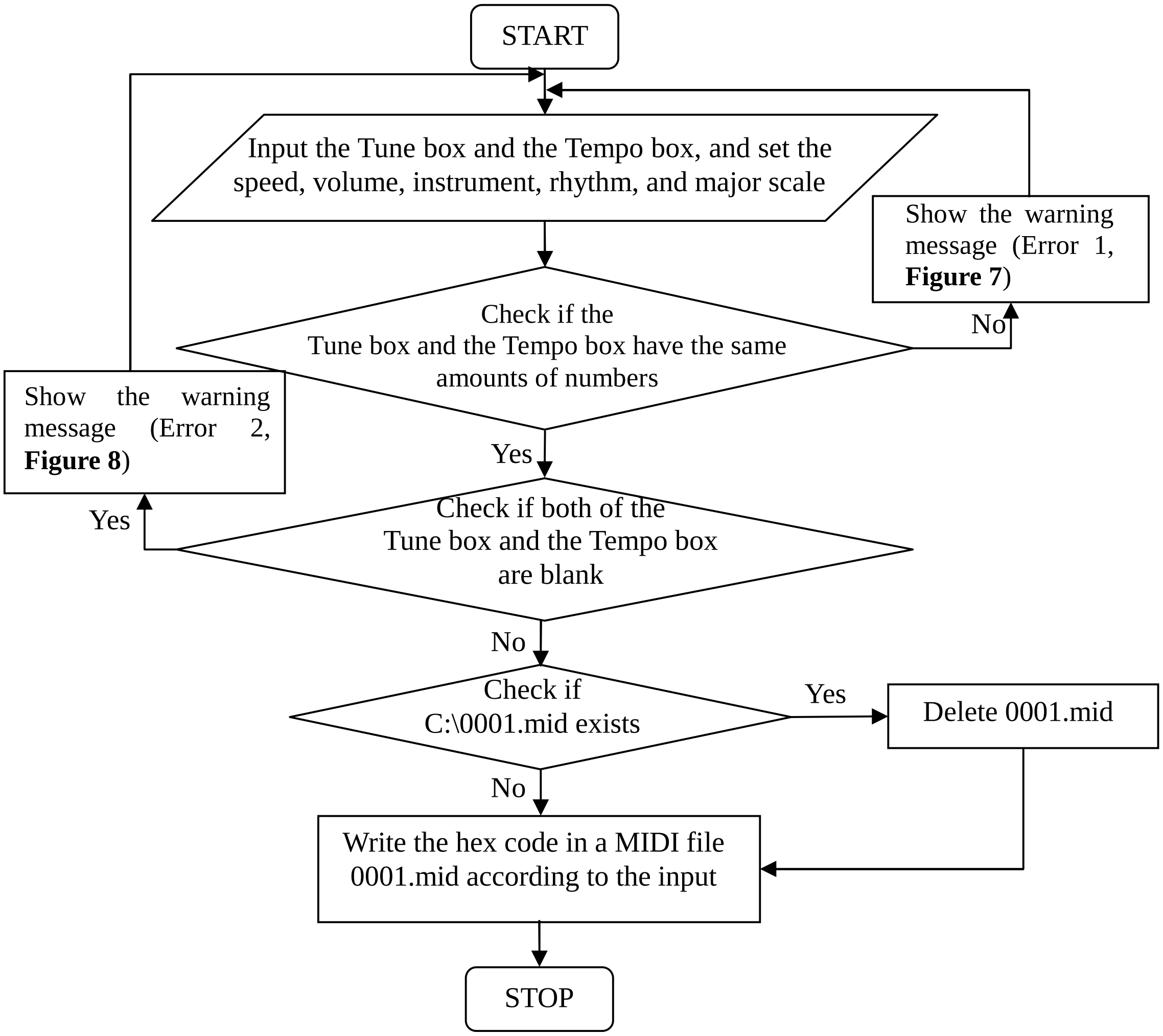

Figure 3. Flowchart of the Visual Basic program.

### 4.1 Data input

This program has 9 kinds of input data, which are tune, tempo, rhythm, speed, tune volume, rhythm volume, instrument, major scale, and repeat. These 9 kinds of input data offer a wide range of options and type styles. Users are allowed to read, edit, copy, cut, and paste the data conveniently.

#### 4.1.1 Tune

In the Tune box, real numbers are used to represent the notes. For instance, “5” represents 5 (Note G), “50” represents $\dot{5}$ (an octave higher than Note G), “500” represents $\ddot{5}$ (two octaves higher than Note G), “-5” represents $\underset{\bullet}{5}$ (an octave lower than Note G), “-50” represents $\underset{\bullet\bullet}{5}$ (two octaves lower than Note G), “5.5” represents $^{\#}5$ (Note $G^{\#}$), and “0” represents the musical rest.

#### 4.1.2 Tempo

Tempos are also digitalized in the Tempo Box. For example, “2” means two beats, and “0.5” means half beat.

#### 4.1.3 Speed

Speed parameter, which ranges from 0 to 10, controls the music playing speed.

#### 4.1.4 Volume

Volume parameters (both tune volume and rhythm volume), which range from 0 to 10, control the sound volume.

#### 4.1.5 Instrument

As shown in **Figure 4**, instrument parameter can control the instrument being used, such as Acoustic Bass, Acoustic Grand Piano, Electric Guitar (jazz), etc.

#### 4.1.6 Major scale

As shown in **Figure 5**, major scale parameter can control the major scale, such as C, $D^b$, D, $E^b$, E, F, $F^{\#}$, G, $A^b$, A, $B^b$, and B.

#### 4.1.7 Rhythm

As shown in **Figure 6**, rhythm parameter can select a rhythm accompaniment, such as Waltz, Rock, Disco, Rumba, etc. If we select *NONE*, then no rhythm accompaniment will be played.

#### 4.1.8 Repeat

This option can specify whether loop playback the melody.

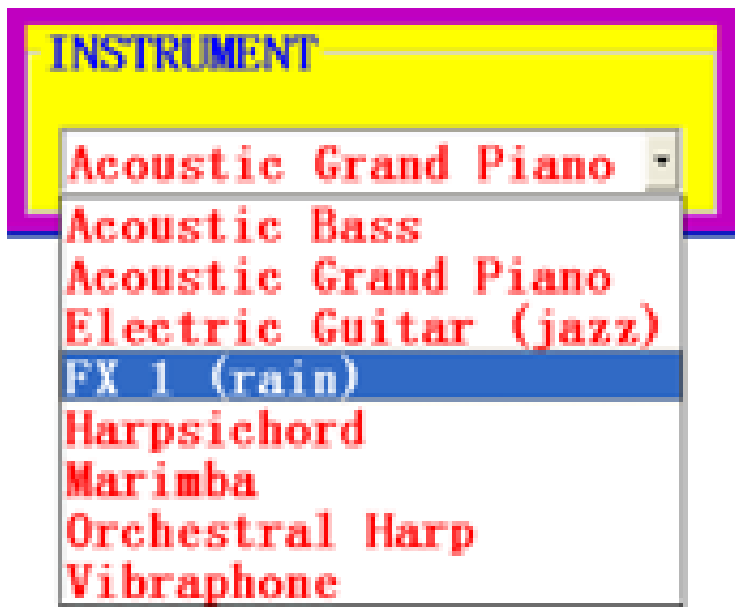

Figure 4. Instrument parameter.

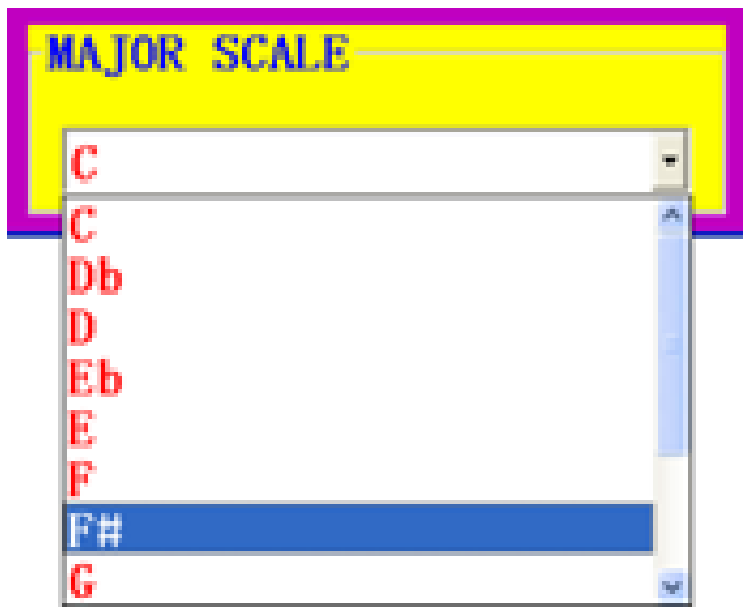

Figure 5. Major scale parameter.

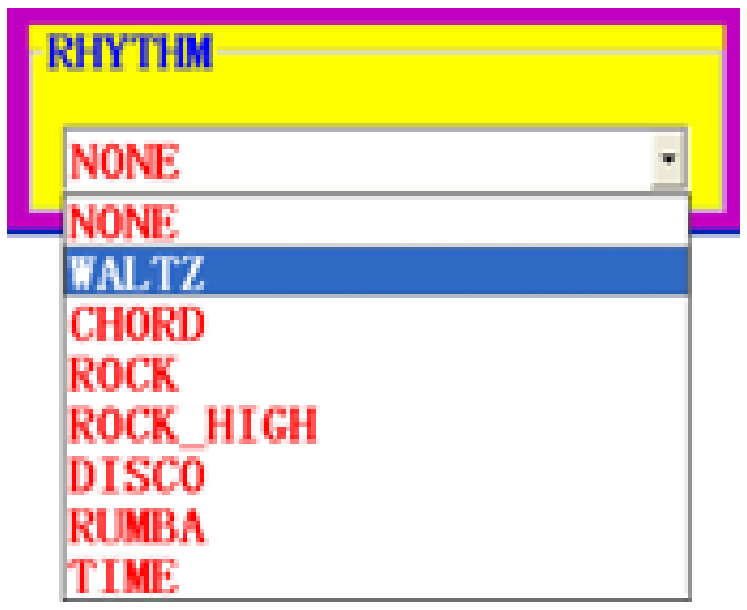

Figure 6. Rhythm parameter.

## 4.2 Data Check

This program can automatically detect and show the quantities of numbers in the Tune box and Tempo box. If they are different, an error message will pop up as shown in **Figure 7**. Similarly, if they are blank, an error message will pop up as shown in **Figure 8**.

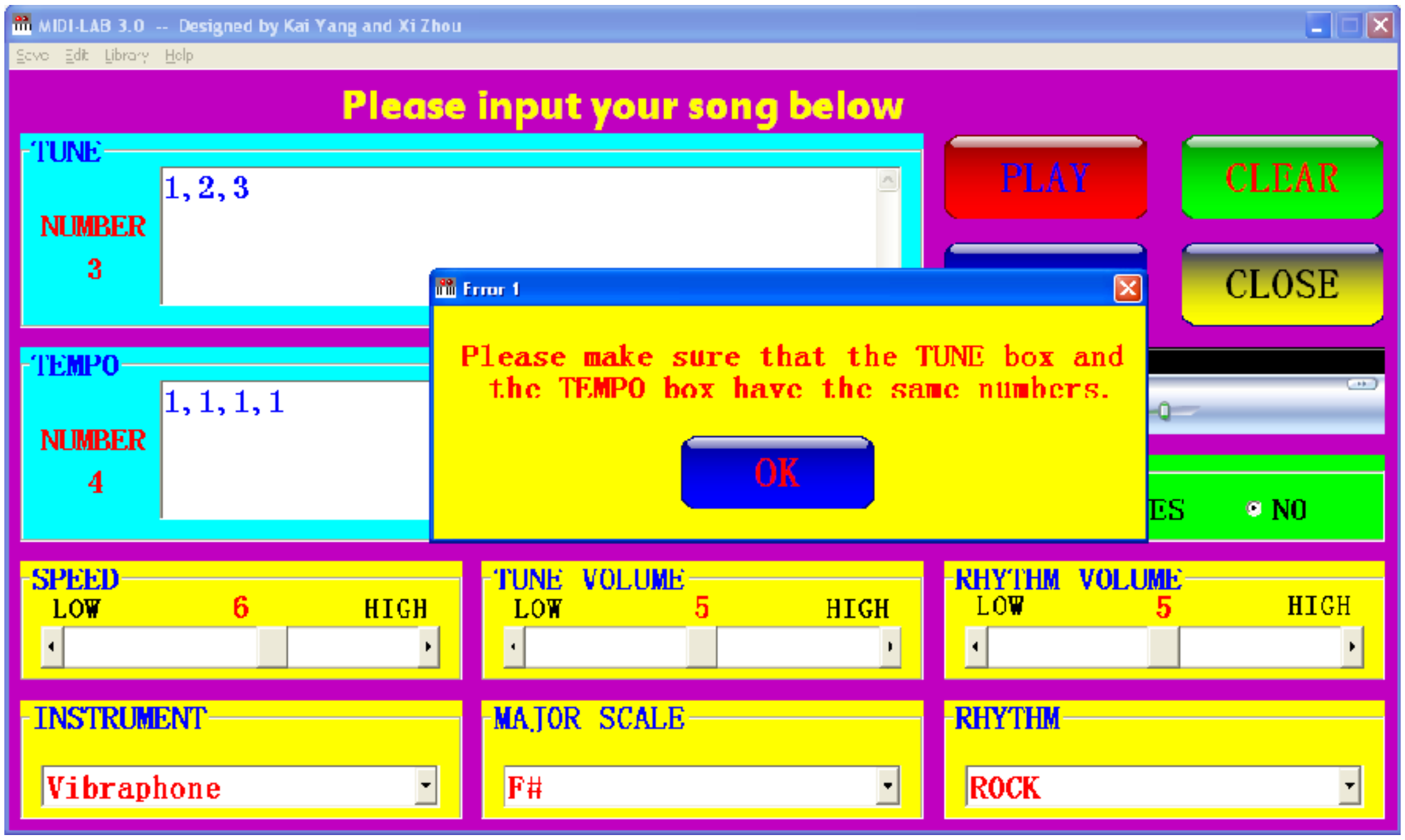

Figure 7. The Error 1 indication.

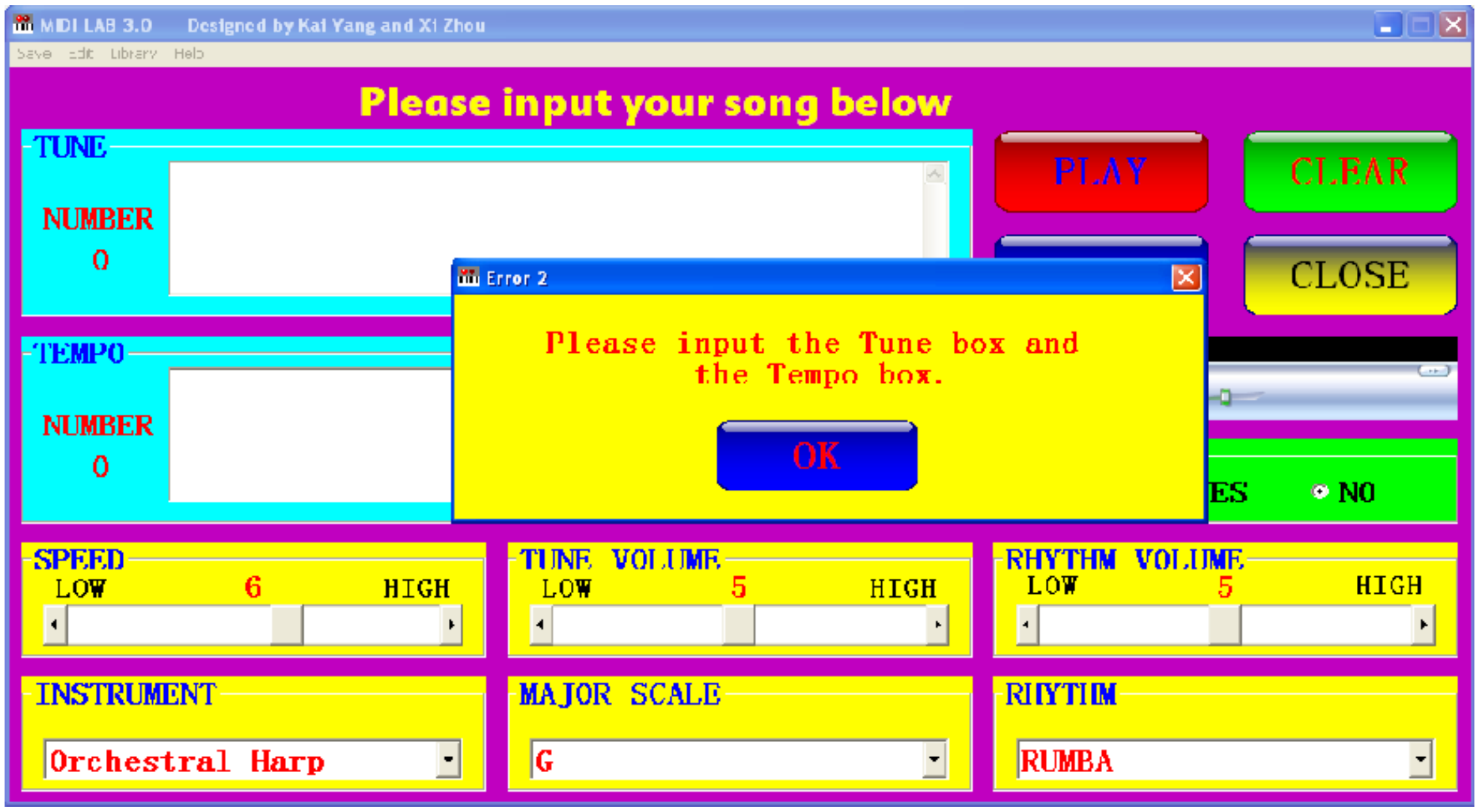

Figure 8. The Error 2 indication.

### 4.3 Data Output

After clicking the PLAY button, this program can automatically write and play a MIDI file 0001.mid according to the input data. Users can also click the Save menu to output the generated MIDI file to a specified directory, as shown in **Figure 9**.

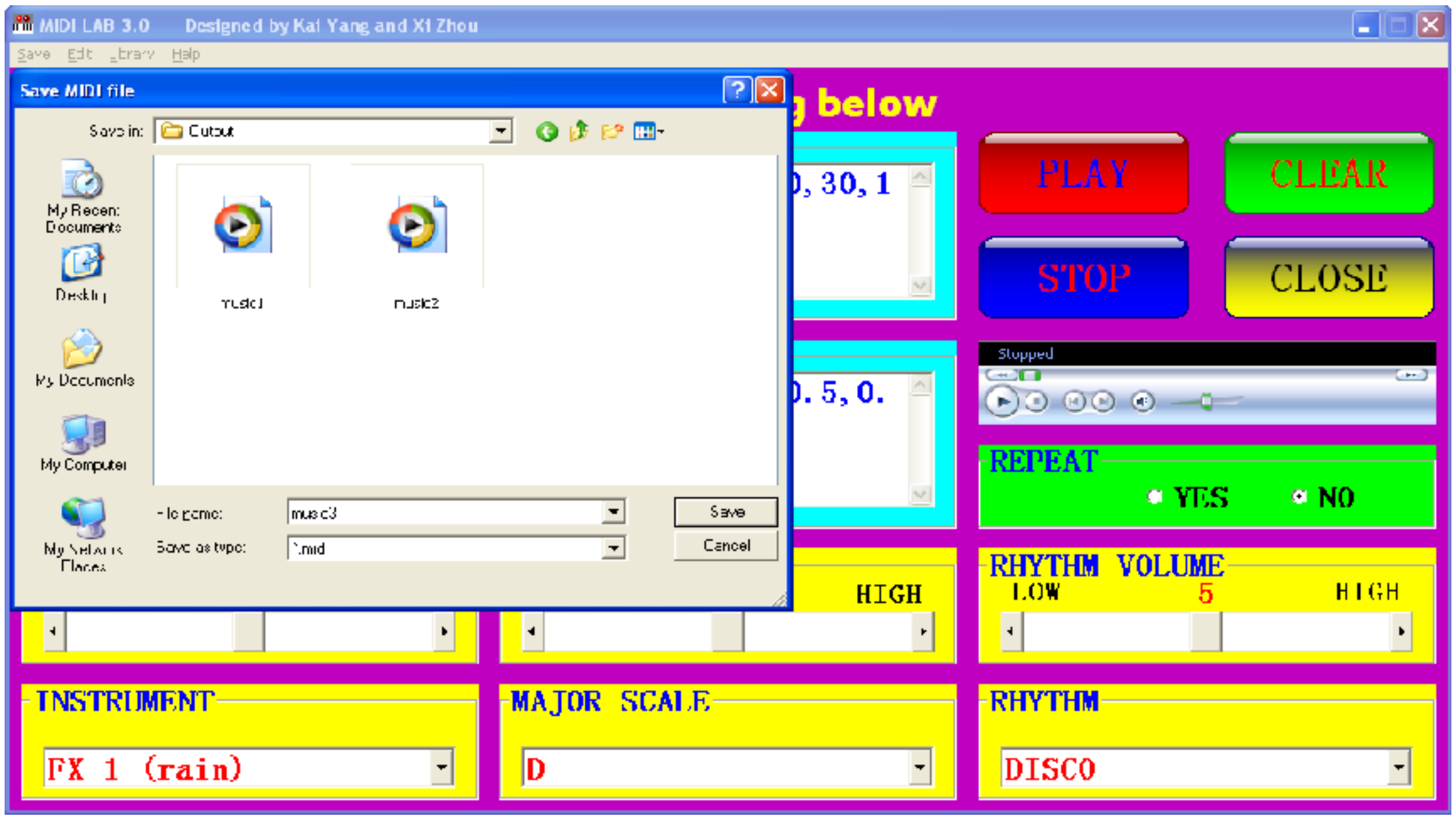

Figure 9. MIDI files output.

# 5 Discussions

## 5.1 Playing chords

Not only can this program play single notes, but it can also play chords of several notes. For example, if we want to play a three-chord in **Figure 10 (a)**, we just need to input the tunes and tempos in **Figure 10 (b)**.

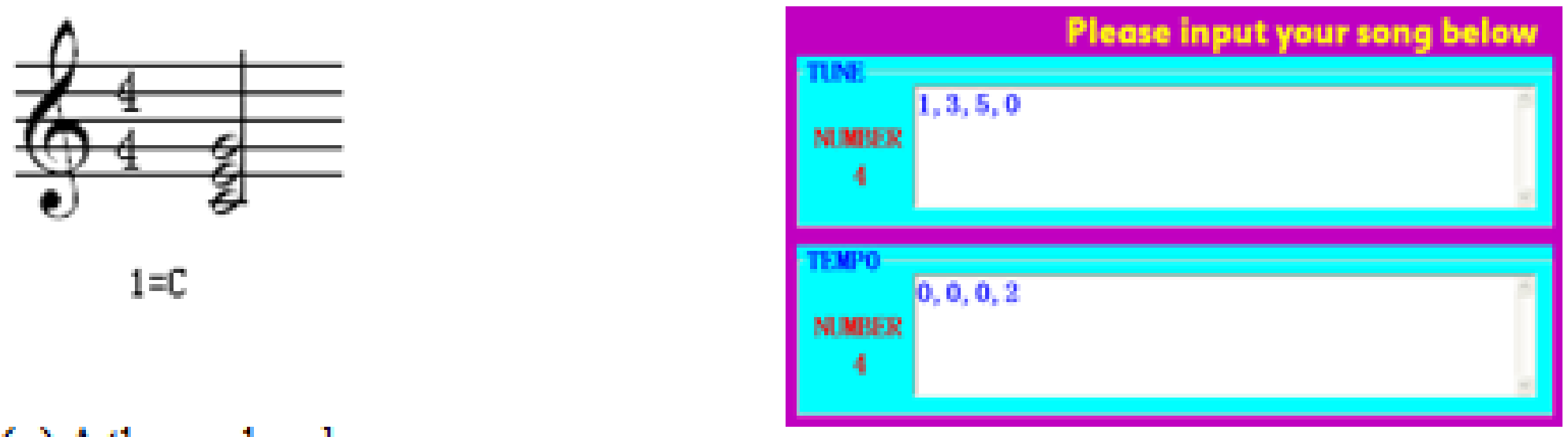

**(a)** A three-chord

**(b)** The input data of a three-chord

Figure 10**.** Example of playing three-chord.

## 5.2 Music library

As illustrated in **Figure 1**, this program, for demonstration purposes, provides a predefined music library, including 4 songs that are “Happy Birthday”, “Christmas Eve Song”, “Amazing Grace”, and “Happy Day”. When the users select a song from the library menu, the corresponding tunes and tempos will be loaded in the boxes automatically.

## 5.3 Parameter options

Users can adjust the parameters, such as speed, volume, instrument, rhythm, and major scale to completely change the character of the music.

## 5.4 Further development

In order to serve more and more customers, MIDI-LAB still needs improvement in the future.

### 5.4.1 Using more track chunks

MIDI-LAB only uses two track trunks in MIDI files (one for melody, the other for rhythm accompaniment). In the future, we will take advantage of several track trunks, so that multiple tunes with different instruments can be played simultaneously.

### 5.4.2 Composing music

Currently MIDI-LAB can only play melodies according to the input parameters. In the future, we will develop an algorithm, so that it can also compose euphonious melodies with different styles, such as Chinese style, Japanese style, Hindu style, etc.

## 6. Conclusions

This paper presents a new powerful and unparalleled Visual Basic program MIDI-LAB for creating MIDI Music. This user-friendly, menu-driven program provides an intelligent human computer interface with digitalized adjustable parameters. Once the users input the desired tunes and tempos, and click the PLAY button, a euphonious melody can be played. With very little practice, the users will also be able to experiment creatively with many different speeds, volume levels, instruments, rhythms and major scales. This program can provide an opportunity to create or change music without the need for additional instruments or players. It can be a valuable tool in many educational, recreational, and commercial applications. MIDI-LAB can also be used to help professional musicians to compose art music. MIDI-LAB will become a practical tool in teaching and reading music, as well as for teaching basic concepts such as pitch and harmony. This could prove a real boon to musically-talented people whose actual technical skills are not yet sufficiently developed to physically perform real-time music. To sum up, MIDI-LAB will provide a creative and encouraging, yet practical, tool for music lovers, teachers, and game makers who need, or wish to create MIDI music.

## 7. ACKNOWLEDGEMENTS

I am heartily thankful to my friend, Prof. K. B. Hom, whose encouragement, guidance and support enabled me to finish this paper.

## Authors

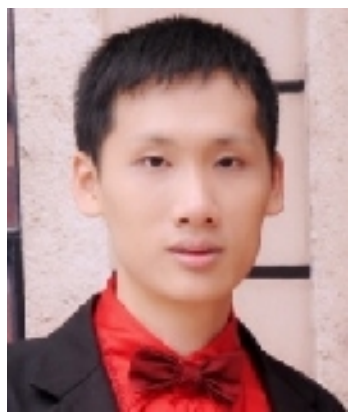

Kai Yang received a bachelor's degree in the Department of Microelectronics from Peking University, China, in 2004, and a Ph.D degree from the Institute of Microelectronics, Chinese Academy of Sciences, China, in 2009. Currently he is pursuing another master's degree on Computational Engineering and Science at University of Utah. He is good at designing, simulating, and fabricating Micro-Electro-Mechanical Systems (MEMS) devices, and developing software programs.

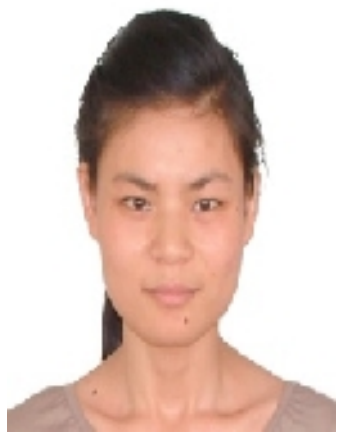

Xi Zhou received a bachelor's degree and continued her graduate study in the Department of Statistics from Capital University of Economics and Business, China, in 2007. Currently she is pursuing her master's degree on Biostatistics at University of Utah. She is good at calculation and programming.